\documentclass[5p]{elsarticle}
\usepackage{amsmath}
\usepackage{bm}
\usepackage{amssymb}
\usepackage{amsfonts}
\usepackage{color}
\usepackage[noautoscale,vcentermath]{youngtab}
\begin{document}

\title{$(2+1)$-$d$ Glueball Spectrum within a Constituent Picture}

\author[rvt,rvt2]{F. Buisseret\corref{cor1}}
\ead[url]{fabien.buisseret@umons.ac.be}
\author[rvt,rvt3]{V. Mathieu}
\ead[url2]{ vincent.mathieu@umons.ac.be}
\author[rvt]{C. Semay}
\ead[url3]{claude.semay@umons.ac.be}

\address[rvt]{Service de Physique Nucl\'{e}aire et Subnucl\'{e}aire,
Universit\'{e} de Mons--UMONS,
Acad\'{e}mie universitaire Wallonie-Bruxelles,\\
Place du Parc 20, B-7000 Mons, Belgium}
\address[rvt2]{Haute \'Ecole Louvain en Hainaut (HELHa), Chauss\'ee de Binche 159, B-7000 Mons, Belgium}
\address[rvt3]{ECT*, Villa Tambosi, I-38123 Villazzano (Trento), Italy }

\date{\today}

\begin{abstract}
The quantum numbers and mass hierarchy of the glueballs observed in $(2+1)$-dimensional lattice QCD with gauge group SU($N_c$) are shown to be in agreement with a constituent picture. The agreement is maintained when going from glueballs to gluelumps, and when the gauge group SO($2N_c$) is taken instead of SU($N_c$).
\end{abstract}



\maketitle

\section{Introduction}

Quantum Chromodynamics (QCD) allows for various types of hadrons. First, the meson and baryon spectra computed from either constituent models \cite{Godfrey:1985xj,Capstick:1986bm} or lattice simulations \cite{Edwards:2011jj,Dudek:2011tt} in $(3+1)$-dimensions were found to be in very good agreement with the experimental spectra \cite{PDG}. Beside these conventional hadrons, QCD does not forbid the existence of other exotic resonances like glueballs, hybrid mesons, tetra- and penta-quarks, etc. 


It is worth commenting a bit the current situation concerning glueballs, since the present works is devoted to that peculiar kind of exotic hadrons. The lightest glueballs are expected to mix significantly with conventional states \cite{Mathieu:2008me}; hence their experimental identification is a challenging task and no firm glueball candidate is known yet \cite{Crede:2008vw}. Constituent and other effective models can in principle drive experimental searches by providing calculations of the glueball mass spectrum and of their preferred decay channels, while lattice QCD, being formulated in Euclidean space-time can only provide the mass spectrum \cite{Morni}. In that sense constituent models are more powerful that numerical simulations, but they rely on the rather strong hypothesis that a relativistic bound state can be approximated by a fixed number of constituents (quark and/or gluons) interacting through an instantaneous potential. 

The main assumption of constituent approaches 
 is readily justified for quarks: The heavy quarks are heavy enough for an expansion in $1/m_Q$ to be relevant \cite{Brambi}, while light quarks are known to develop an effective mass $m\sim300$ MeV at low energy \cite{qmass}. The situation is different for glueballs. The gluon mass generation was postulated long time ago \cite{Cornwall:1981zr} but received support from numerical simulations only recently \cite{Bogolubsky:2007ud}. After \cite{Cornwall:1981zr}, it has been claimed that glueballs could be described by a constituent picture thanks to this dynamically generated gluon mass \cite{Cornwall:1982zn,Szczepaniak:1995cw}. In the $(3+1)$-dimensional case, it turned out that the constituent model spectrum matches the lattice one provided that the glueball wave function is built from the transverse degrees of freedom  of the constituent gluons only \cite{Szczepaniak:1995cw,Mathieu:2008bf, Mathieu:2009cc,Mathieu:2008pb}. The gluonic longitudinal component inherent to massive vector particles actually decouples for physical observables thanks to current conservation \cite{Aguilar:2011xe}.

Because of the obvious connection with experimental data, glueball properties have mostly been studied within a $(3+1)$-dimensional spacetime. However, some results are known in $(2+1)$-dimensions, a case that is particularly relevant mostly for two reasons. First, decreasing the number of spatial dimensions can significantly reduce the computational effort in lattice calculations and serve as a toy model for the $(3+1)$-dimensional theory. Second, the spectrum of a Yang-Mills gauge theory with two spatial dimensions exhibits peculiar features, that will be detailed below, and whose understanding is intrinsically challenging. The glueball spectrum in $(2+1)$-dimensions has been computed in \cite{Teper,Tepnew} for the gauge group SU($N_c$), and the lowest-lying states have been studied with SO($2N_c$) in \cite{Teper2}. 

To our knowledge, the structure of the $(2+1)$-$d$ glueball spectrum has not been studied yet by resorting to constituent approaches. As shown by lattice computations, the gluon propagator reaches a finite value in the infrared in $(2+1)$-$d$ as well \cite{Cucchieri:2003di}. A constituent model should then be also appropriate to describe the glueball spectrum obtained in numerical simulations. This study aims precisely at this and is organized as follows: The basic features of a constituent picture are presented in Sec.~\ref{sec:cons} with the numerical technique used to solve the eigenequations associated. The two- , three- and four-gluon bound states are then studied in the SU($N_c$) case in Secs.~\ref{sec:2}, \ref{sec:3} and \ref{sec:4} respectively. The extension of our formalism to arbitrary gauge groups is presented in Sec.~\ref{sec:gauge} while gluelumps are discussed in Sec.~\ref{sec:1}. Concluding comments are given in Sec.~\ref{sec:conclu}.

\section{Constituent picture}\label{sec:cons}

In $(D+1)$ dimensions, the gauge field has $D-1$ independent components because of gauge invariance and transversality. In $D=2$, only one component is left. So, at the level of our model, a constituent gluon will be seen as a ``scalar" particle in the adjoint representation of the gauge group, with an intrinsic parity equal to $-1$. 

As explained in the introduction, gluon mass generation suggests that a constituent description of glueball states may be relevant. Any glueball state can then be interpreted as a bound state of a given number ($n_g\geq2$) of interacting  constituent gluons. The interaction being confining, one should in a first approximation be able to label the various states by a harmonic oscillator band number related to some Jacobi coordinates:
\begin{equation}
N=K+n_g-1, \quad {\rm with}\quad K=\sum^{n_g-1}_{i=1} (2n_i+\vert j_i\vert )=2n+J,
\end{equation}
and assume that, for the lightest states at least, the mass $M$ is $\propto n_g$. In the above definition, $j_k$ is the eigenvalue of the two-dimensional angular momentum related to the $k$th Jacobi coordinate, namely  $\hat j_k=-i(x_k\partial_{y_k}-y_k\partial_{x_k})$. Physical states are then characterized by a spin $J\geq 0$ such that  $\hat J^2 \vert J\rangle=J^2\vert J \rangle$, where $\hat J=\sum^{n_g-1}_{i=1} \hat j_i$.

The parity $\hat P$ is defined as the operation $(x,y)\rightarrow (x,-y)$ \cite{Teper}: It is crucial to note that $\hat P\vert J\rangle =\vert - J\rangle$. Therefore, the state $\vert J^P\rangle=\vert J\rangle + P \vert -J\rangle$ has the parity $P$, with $J\neq 0$. The case $J=0$ needs to be separately considered. Under the reasonable assumption that the Hamiltonian of the theory commutes with the parity operator, the states  $\vert (J>0)^\pm\rangle$ have the same mass. This is a characteristic signature of $(2+1)$-dimensional quantum mechanics.  

Note that,  in $(2+1)$-$d$, it is in general convenient to use the complex coordinates $ x_\pm=x\pm i  y$ instead of the Cartesian ones. The spin operator then reads $\hat J= x_+ \partial_+ - x_-\partial_-$ ($\partial_{\pm}=\partial_{x_{\pm}}$), and the parity acts as a complex conjugation, \textit{i.e.} $\hat P x_\pm=x_\mp$.

In the following, we will need to solve two- and three-body relativistic $(2+1)$-$d$ eigenequations. A full numerical three-body calculation in this case is out of the scope of this exploratory work, but one can nevertheless resort to a semiclassical mass formula in order to get a correct estimate of the mass spectrum. We will use an extended version of the Dominantly Orbital State (DOS) method developed at the origin for two-body $(3+1)$-$d$ hadronic systems \cite{olss97,doscs}. It can be generalized to treat three-body system with an arbitrary kinetic part for any number of dimensions \cite{dos}. The idea is to quantize the radial excitations around classical circular orbits with quantized angular momentum. In principle, accurate results can only be computed for small radial excitations and high orbital excitation, but it can be shown that very good results can be obtained for the whole spectra in the cases of two- and three-body relativistic systems with linear interactions \cite{olss97,doscs,dos}. A particularity for $(2+1)$-$d$ systems is the impossibility to use the DOS method to treat states with a vanishing total orbital momentum, but then a WKB computation can be used instead \cite{dos}. 

\section{Two gluons}\label{sec:2}

The charge conjugation of a color-singlet bound state of two constituent gluons is always positive: A color wave function like $\delta_{ab} A^a_\mu A^b_\nu$ is such that ${\cal C} (\delta_{ab} A^a_\mu A^b_\nu) = + \delta_{ab} A^a_\mu A^b_\nu$. Note that this is valid whatever the gauge group since one can always construct the invariant tensor $\delta_{ab}$. 

To establish the allowed $J^P$, we recall that the various glueball wave functions can be expressed in a harmonic oscillator basis, for which analytical computations are particularly straightforward. Let ${\bm \rho}={\bm x_1}- {\bm x_2}$ be the relative (spatial) coordinate between the two constituent gluons. Then, in a harmonic oscillator basis, the spatial wave function of any two gluon state has the form $\psi_{gg;J^{PC}}=f_J(\rho_+,\rho_-)\,  {\rm e}^{-\beta \rho_+\rho_-}$, where $f(\rho_+,\rho_-)$ is a polynomial in $\rho_\pm$, whose degree is related to the energy level of the considered state.  

The spin is given by the eigenvalue of $\hat J= \rho_+ \partial_+ - \rho_-\partial_-$, so $J=0$ states must be functions of the product $\rho_+\rho_-$ only. Since $\hat P$ interchanges $\rho_+$ and $\rho_-$, the parity of these scalar states is positive. The Pauli principle has finally to be imposed and the state must be symmetrized -- constituent gluons are indeed bosons. By definition, $P_{12} \rho_\pm=\rho_\mp$, so the scalar states are symmetric. One sees that $f_{0^{++}}(\rho_+,\rho_-)=(\rho_+\rho_-)^n+\dots$ generates a tower of $0^{++}$ glueballs at the energy level $N=2n+1$. In the following, $f_{J^{PC}}$ will always denote the term with the highest degree in the corresponding polynomial and the dots, denoting the terms with lower degrees, will be dropped.

Higher spin states can be built by using $f_{(J>0)^{P+}}=(\rho_+^J+P \rho_-^J)(\rho_+\rho_-)^{n}$, generating a glueball at energy level $N=2n+J+1$. However, only even values of $J$ ensure a totally symmetric state: The allowed two-gluon glueball quantum numbers are finally $0^{++}$, $2^{\pm +}$, $4^{\pm +}$, \dots As observed on the lattice \cite{Teper}, the scalar and tensor glueballs are indeed the lightest states, which is natural in view of the present approach: They are made of a minimal number of constituent gluons. As observed in Table~\ref{tab}, the mass hierarchy of the lightest sates observed on the lattice fits in the classification developed above. The harmonic oscillator degeneracy is broken since the underlying dynamical model is not expected to be a nonrelativistic harmonic oscillator. We remark that no $0^{-+}$ two-gluon bound state can be formed ; this was not the case in $(3+1)$-$d$. The pseudoscalar glueball is thus expected to be well heavier than the scalar one, as observed in Table \ref{tab2}.  

Let us first compare our approach to the large-$N_c$ limit of the lattice results since the coupling between states with different number of gluons is expected to vanish at large $N_c$ \cite{Liu}. The simplest model able to describe a bound state of two constituent gluons is the Hamiltonian
\begin{equation}\label{H2B}
H=2\sqrt{{\bm p^2}+m^2_g}+\sigma_{adj} \rho ,
\end{equation}
where we allow for a gluon mass $m_g$ and where we assume the Casimir scaling of the string tension,
\begin{equation}\label{sadj}
\sigma_{adj}=\frac{C_g}{C_q}\sigma .
\end{equation}
In the above equation, $C_g$ ($C_q$) is the quadratic Casimir operator for the gluon (quark) and $\sigma$ is the fundamental string tension. The one-gluon-exchange potential has been neglected: In $(2+1)$-$d$, it is proportional to $\ln(\nu \rho)$, where $\nu$ is some constant. Such a potential being weakly confining, its influence is expected to be similar to that of the linearly confining term. So it should not be as crucial as the Coulomb form in $(3+1)$-$d$. The large-$N_c$ limit of our model reads 
\begin{equation}\label{Hlnc}
{\cal H}=\lim_{N_c\rightarrow\infty}\frac{H}{\sqrt\sigma}=2\sqrt{{\bm q^2}+\mu^2}+2\, r,
\end{equation}
where $\bm q$ and $\bm r$ are dimensionless conjugate variables. The free parameter is $\mu=\lim_{N_c\rightarrow\infty} m_g/\sqrt{\sigma}$, that we set equal to $\mu=0.7$. 
The agreement between our model and the lattice data of \cite{Teper} can be appraised in Table \ref{tab}: It is very satisfactory with respect to the simplicity of our approach. As mentioned before, the numerical results have been obtained thanks the DOS method \cite{olss97,doscs,dos}. They have been checked by the use of Lagrange-mesh method \cite{LMM} which can be easily extended to $(2+1)$-$d$ systems while preserving its high accuracy. 

 
A fit of the mass spectrum in the large-$N_c$ limit shows that the masses of the $n=0$ states, that have $J^{PC}=({\rm even})^{++}$ quantum numbers, are located on the Regge trajectory $J=\alpha' M^2+\alpha_0$, with $ \sigma\alpha'= 0.0625$ and $\alpha_0=-0.976$. This compares favorably with the lattice result  $ \sigma\alpha'= 0.0611(25)$, $\alpha_0=-1.144(71)$, found in \cite{Meyer04}. Such a Regge trajectory may be seen as the two-dimensional equivalent of the Pomeron.

 
\section{Three gluons}\label{sec:3}

With any gauge group, a totally antisymmetric color singlet can be made from three constituent gluons, which has a positive charge conjugation: It is realized through a coupling involving the structure constants $f^{abc}$. In the special case of SU($N_c>2$), the totally symmetric invariant tensors $d^{abc}$ can be used to build a totally symmetric color singlet from three constituent gluons, with a negative charge conjugation \cite{gluboul}. The Pauli principle demands the spatial wave function to be totally (anti)symmetrized for $C=(+)-$ three-gluon glueballs. If the problem of writing correctly symmetrized quantum three-gluon states is not fully solved yet in $(3+1)$-$d$ (a formal procedure can be found in \cite{gieb}), the reduction of gluonic degrees of freedom will allow an explicit determination of these states in $(2+1)$-$d$. 

We closely follow here Ref. \cite{Murt}, where a straightforward method to generate two-dimensional harmonic oscillator states is presented and applied to explicitly build all states with a well-defined permutation symmetry up to the $K=4$ band in our notation. One first defines the Jacobi coordinates ${\bm \rho}=({\bm x}_1-{\bm x_2})/\sqrt 2$, ${\bm \lambda}=({\bm x}_1+{\bm x_2}-2{\bm x_3})/\sqrt 6$, and then works with the coordinates $u_\pm=\left\lbrace \rho_\pm, \lambda_\pm\right\rbrace $. This is convenient because the action of the parity operator and of the permutation operators on these four coordinates is straightforwardly computed \cite{Murt}.

In a harmonic oscillator basis, the spatial wave function of any three gluon state has the form $\psi_{ggg;J^P}=f_{J^P}(\rho_\pm,\lambda_\pm)\,  {\rm e}^{-\beta (\rho_+\rho_-+\lambda_+\lambda_-)}$, where $f_{J^P}$ is a polynomial in $\left\lbrace \rho_\pm,\lambda_\pm\right\rbrace $. Totally symmetric or totally antisymmetric polynomials can then be built and, by linear combination, one can obtain polynomials that are also eigenstates of $\hat j^2=(\rho_+\partial_+-\rho_-\partial_-+\lambda_+\partial_+-\lambda_-\partial_-)^2$ and of the parity ($u_\pm \partial_\pm=u_\pm \partial_{u_\pm}$). For the parity, one has to recall that the intrinsic negative parity of the constituent gluons adds a global minus sign to the result. 

The terms with the highest degree of the functions $f_{J^P}$ leading to given glueball states are listed in Table \ref{tab2}. As expected, selection rules arise because of symmetrization, and the allowed quantum numbers we find perfectly match those found on the lattice. A first observation is that the lightest state is a $0^{--}$ one,  such that $M_{0^{--}}/M_{0^{++}}=1.45\approx 3/2$, in agreement with a constituent gluon picture. A second observation is that a $0^{-+}$ state is at first present in the $K=6$ band: This is an indication that the splitting between the scalar and pseudoscalar states has to be large, as observed on the lattice. A last observation is that no $0^{+-}$ state can be formed with three constituent gluons. Indeed, the negative charge conjugation asks for a totally symmetric state for color and space. As can be observed from \cite{Murt}, any $0^{P-}$ three-body state must have a negative parity. So this glueball is at least a four-gluon one. Notice that $M_{0^{+-}}/M_{0^{++}}=2.32\gtrsim 2$, as suggested by a constituent gluon picture. 

Our two-body model can be generalized to a three-body case. In a flux-tube picture, the static potential describing three adjoint sources has a $\Delta$-shape as far as SU($N_c$) is chosen \cite{bicu}: The gluons are linked by fundamental strings, and the Hamiltonian reads $H=\sum^3_{i=1}\sqrt{\bm p^2_i+m^2_g}+\sum^3_{i<j=1}\sigma |{\bm x}_i -   {\bm x}_j |$, whose large-$N_c$ limit is 
\begin{equation}\label{H3B}
{\cal H}=\lim_{N_c\rightarrow\infty}\frac{H}{\sqrt\sigma}=\sum^3_{i=1}\sqrt{\bm q^2_i+\mu^2}+\sum^3_{i<j=1} |{\bm r}_i -   {\bm r}_j |,
\end{equation}
where $\bm q_i$ and $\bm r_i$ are dimensionless conjugate variables. 

The mass spectrum of Hamiltonian~(\ref{H3B}) is given in Table~\ref{tab2}. Masses of three-gluon glueballs have been computed with the DOS method. The mass of the $0^{--}$ states can also be straightforwardly obtained by performing variational computations with Gaussian trial wave functions of the type ${\rm e}^{-\beta (\rho^2+\lambda^2)}\times (\textrm{polynomial})$. Applying techniques similar to those of \cite{mathgauss} in the case of two spatial dimensions, one obtains for example the upper bound $M_{0^{--}}/\sqrt\sigma=6.49$, with a size $\beta=1.03$. This value is actually close to our semiclassical estimates, as are the masses found for the $0^{--*}$ and $0^{--**}$ glueballs. This is another check of the validity of the DOS approximation scheme. 

It can be observed that, although a quantitative agreement is not reached, the trend of the spectrum is correctly reproduced. Remark that the $0^{-+}$ glueball is the only state for which our estimate is clearly incompatible with the lattice result. Its overestimation might be the result of a lack of precision in the prediction of the three-body radial slope, which affects more severely this $n=3$ state.  

The masses of the $n=0$ eigenstates of (\ref{H3B}) with $J^{PC}=({\rm odd})^{--}$ are located on the Regge trajectory $J=\alpha' M^2+\alpha_0$, with $\sigma\alpha'\approx 0.0481$ and $\alpha_0=-1.964$. Such a low intercept, together with a slope similar to but lower than the two-gluon case, has also been pointed out in \cite{Meyer04} .
 
\section{Four gluons}\label{sec:4}

As previously said, the $0^{+-}$ glueball has to be made of at least four gluons. To estimate its mass, we follow a procedure similar to what has been done in \cite{gluboul}. According to the results of this last reference (that have been obtained for SU(3) but that are valid for SU($N_c$) too), a color singlet four gluon state with $C=-$ should have a spatial symmetry of the form $\scriptsize{\young(13,24)}$. There are thus two totally antisymmetric two-gluon clusters: This can be reached by a $J=1$ two-gluon wave function of the form $\rho_+ \pm \rho_-$. In the flux tube picture, where two fundamental strings originate from each constituent gluon, the mass of such a cluster is equal to $4.22$ in units of $\sqrt\sigma$.   

The quantum state of the clusters being known, we can couple them in order to
obtain a $0^{+-}$ state. In order to satisfy at best the mixed symmetry of the four-gluon state, we ask for the two clusters to be identical and for the two-cluster state to be the lightest $J=0$ state. The glueball mass computed this way is 10.11 in units of $\sqrt\sigma$: Although being a seemingly rough estimate, it is in good agreement with the lattice result 9.47(116) \cite{Teper}. 

\section{Arbitrary gauge group}\label{sec:gauge}

The behavior of thelightest glueballs with respect to a change of gauge group can be straightforwardly found. Up to a rescaling of the relative coordinates, Hamiltonian (\ref{H2B}) can be written as ${\cal H}=\sqrt\kappa\left[  2\sqrt{\bm q^2+(\mu/\sqrt\kappa)^2}+, r\right] $, with $\kappa=C_g/C_q$. According to the pinch technique study \cite{Agui}, nonperturbative effects generate a nonzero gluon mass in $(2+1)$-$d$ Yang-Mills theory, which is computed from the zero-momentum limit of the gluon propagator. A fit to lattice results in Landau gauge leads to $m_g\approx 0.146\, C_g \, g^2$ \cite{Agui}.  Moreover, in $(2+1)$-$d$, one can compute that the string tension is given by $\sqrt\sigma=g^2 \sqrt{C_g C_q/4\pi}$ \cite{nair}. Combining both results leads to $m_g\approx 0.146 \sqrt{ 4\pi\kappa\sigma}$, or $\mu/\sqrt\kappa\approx 0.146 \sqrt{4\pi}\approx 0.52$. Note that, in the large-$N_c$ limit, $\kappa=2$ and we get $\mu\approx 0.73$, a value perfectly matching the one we used to reproduce at best the low-lying glueball spectrum. 

From this discussion we can conclude that $\mu/\sqrt\kappa$ should be a gauge-group independent quantity and that the masses of the two-gluon glueballs should obey a simple scaling law. In particular, we expect
\begin{equation}
\frac{M_{gg}}{\sqrt\sigma}=\sqrt\frac{C_g}{C_q} A,
\end{equation}
with $A$ a gauge-group-independent constant that we compute to be equal to $A=2.87$ for the lightest $0^{++}$ glueball. Equivalently, we expect the glueball masses normalized to $\sqrt{\sigma_{adj}}$ to be gauge-group independent. 

\begin{figure}[h]
\begin{center}
\includegraphics*[width=0.45\textwidth]{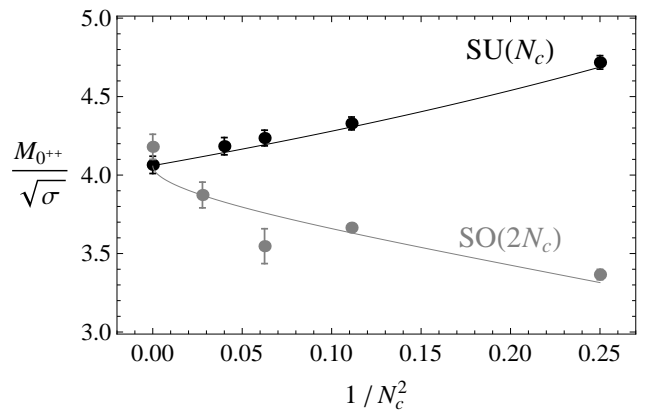}
\caption{Evolution of the $0^{++}$ glueball mass versus $N_c$ for the gauge groups SU($N_c$) (in black) and SO($2N_c$) (in gray). Our model (solid lines) is compared to the lattice data (points) of \cite{Teper} for SU($N_c$) and of \cite{Teper2} for SO($2N_c$).
}
\label{fig}
\end{center}
\end{figure}

The recent lattice work \cite{Teper2} provides us a way to check this last equation since the behavior of the scalar glueball in $(2+1)$-$d$ versus $N_c$ has been computed for both SU($N_c$) and SO($2N_c$). Knowing that $\kappa=2N_c^2/(N_c^2-1)$ for SU($N_c$) and $2(N_c-2)/(N_c-1)$ for SO($2N_c$), one can plot the evolution of the scalar glueball mass versus $N_c$. This is done in Fig. \ref{fig}, where it can be seen that the lattice behavior is well reproduced by our scaling argument.

The masses of some $J^{PC}=({\rm even})^{++}$ glueballs, including highly excited ones, have been computed in $(2+1)$-$d$ lattice QCD with gauge group SU(2) in \cite{Tepnew}. The proposed scaling rule allows to compute the corresponding masses within our model; they compare very well with the lattice data as shown in Table \ref{tab}.

\section{Gluelumps}\label{sec:1}

Gluelumps are bound states of the gluonic field when a static, color adjoint, source, is added. Only one constituent gluon is then necessary to form a color singlet. 
In this case, the Hamiltonian is ${\cal H}=\sqrt{{\bm q^2}+\mu^2}+\kappa\, r$, whose lightest eigenstate (a $0^{--}$ one) has a mass equal to $M_g/\sqrt\sigma=3.11$ for SU(2), still with $\mu=0.7$, while we find the lightest glueball to have a mass $M_{gg}/\sqrt\sigma=4.69$ for SU(2). This can be compared to the lattice study \cite{Philip}, where is has been found that, in SU(2) computations, the link between the adjoint string breaking length scale and the lightest glueball mass is $M_{gg} r_b=10.3\pm1.5$. However, the adjoint string breaking energy is expected to be twice the lightest gluelump mass: $\sigma_{adj} r_b=2 M_g$. Using (\ref{sadj}), the result of \cite{Philip} can be rewritten as
\begin{equation}
\frac{M_g}{\sqrt\sigma}\frac{M_{gg}}{\sqrt\sigma}=\frac{4}{3} (10.3\pm 1.5) =13.7\pm 2.0.
\end{equation}
We find 14.6 for the l.h.s.\ of this relation, which is in good agreement with the lattice results.

\section{Conclusion}\label{sec:conclu}

We have analyzed the structure of the glueball spectrum in $(2+1)$-$d$ by resorting to a constituent picture: Each glueball is described as a bound state of a given number of transverse, massive, constituent gluons. Pauli's principle leads to selection rules: The allowed $J^{PC}$ states match the corresponding lattice results \cite{Teper}. 

\begin{figure}[h!]
\begin{center}
\includegraphics*[width=0.45\textwidth]{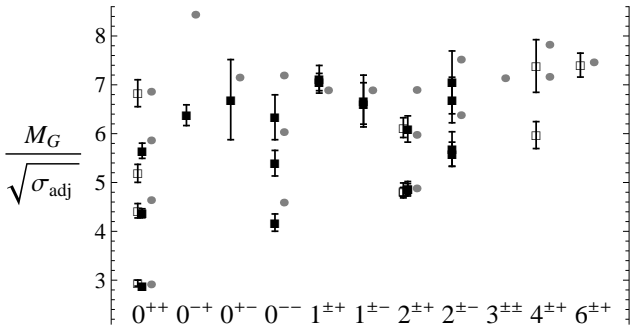}
\caption{Glueball masses in units of $\sqrt{\sigma_{adj}}$, given by (\ref{sadj}), for some $J^{PC}$. All data are taken from Tables~\ref{tab} and \ref{tab2}; lattice data from \cite{Teper} (filled squares) and \cite{Tepnew} (empty square) are compared to our model's results (gray dot).
}
\label{fig2}
\end{center}
\end{figure}

A simple dynamical model -- relativistic kinematics plus linear confinement -- leads to a satisfactory, although not fully quantitative, agreement with the glueball and glue\- lump masses found in \cite{Teper,Meyer04,Philip}. A summary plot is given in Fig.~\ref{fig2}. Our model leads moreover to a simple prediction concerning the modification of the glueball masses with respect to a change of gauge group. This prediction compares favorably with the recent data \cite{Teper2}.  

A useful but quite technical task that we leave for future work will be a more accurate computation of the mass spectrum by using \textit{e.g.} a large harmonic oscillator basis or a large Gaussian expansion. Finally, we remark that a lattice computation of the glueball spectrum with other gauge groups than SU($N_c$) would be very interesting in view of checking further the predictions of our model. 

\section*{Acknowledgements}

C.S. thanks Gwendolyn Lacroix for a useful suggestion.

\begin{table}[ht]
\caption{Glueball masses (in units of $\sqrt\sigma$)  classified by band number $K$ for the different allowed $J^{PC}$ in two-gluon states . The highest degree term, $f_{J^{PC}}$, in the harmonic oscillator wave functions is shown, together with the large-$N_c$ lattice results \cite{Teper} or the SU(2) results \cite{Tepnew}  when available. Results of our model are shown, first computed with the DOS approximation \cite{dos}, then with the Lagrange mesh method  (LMM) \cite{LMM}. }
\begin{tabular}{cllcccccc}
\hline\hline
 $K$ & $J^{PC} $ & $f_{J^P}$ & $M_G/\sqrt\sigma$ & \multicolumn{2}{c}{Model} & $M_G/\sqrt\sigma$ & \multicolumn{2}{c}{Model}  \\
  & & & for $N_c\rightarrow\infty$ \cite{Teper} & DOS & LMM & for SU(2) \cite{Tepnew} & DOS & LMM \\
\hline 
 0 & $0^{++}$ & 1  & 4.065(55) & 4.12 & 4.06  & 4.80(10) & 4.76 & 4.69 \\
   2 & $0^{++*}$ & $\rho_+\rho_-$  &6.18(13) &  6.56 & 6.53 & 7.22(24) & 7.57 & 7.54\\
        & $2^{++ }$ & $\rho_+^2+ \rho_-^2$  &6.88(16) & 6.90 & 6.91 & 7.85(15) & 7.97 & 7.98\\
        & $2^{-+ }$ & $\rho_+^2- \rho_-^2$  &6.89(21) &  6.90 & 6.91 & & 7.97 & 7.98\\
     4  & $0^{++**}$ &  $(\rho_+\rho_-)^2$  &7.99(22) & 8.29 & 8.27  & 8.47(30) & 9.57 & 9.55 \\
        & $2^{++*}$ &  $(\rho_+^2+ \rho_-^2)(\rho_+\rho_-)$  & & 8.45 &  8.50 & 7.90(25) & 9.76 & 9.81 \\
        & $2^{-+*}$ &  $(\rho_+^2- \rho_-^2)(\rho_+\rho_-)$  &8.62(38) &  8.45 & 8.50 & & 9.76 & 9.81\\
         & $4^{++}$ &  $\rho_+^4+ \rho_-^4$  & &  8.91 & 8.92  & 9.75(45) & 10.29 & 10.30\\
          & $4^{-+}$ &$\rho_+^4- \rho_-^4$  &   &  8.91 & 8.92 & & 10.29 & 10.30 \\
    6 & $0^{++***}$ &   $(\rho_+\rho_-)^3$ & &9.70 & 9.67 &11.15(45) & 11.20 & 11.17 \\ 
       & $2^{++**}$ & $(\rho_+^2+ \rho_-^2)(\rho_+\rho_-)^2$ & & 9.75& 9.85& 10.00(33) & 11.26 & 11.37 \\
        & $2^{-+**}$ & $(\rho_+^2- \rho_-^2)(\rho_+\rho_-)^2$ & & 9.75& 9.85&  & 11.26 & 11.37 \\
          & $4^{++*}$ &  $(\rho_+^4+ \rho_-^4)(\rho_+\rho_-)$  & &10.13   &10.17  & 12.06(88) & 11.70 & 11.74 \\
          & $4^{-+*}$ &$(\rho_+^4- \rho_-^4)(\rho_+\rho_-)$  &   &  10.13 &10.17  & & 11.70 & 11.74 \\
          & $6^{++}$ &  $\rho_+^6+ \rho_-^6$  & &  10.55 & 10.56 & 12.09(40) &12.18 & 12.19 \\
          & $6^{-+}$ &$\rho_+^6- \rho_-^6$  &   & 10.55 & 10.56 & &12.18 & 12.19  \\
\hline\hline
\end{tabular}
\label{tab}
\end{table}

\begin{table}[h]
\caption{Same as Table \ref{tab} for three-gluon glueballs. Computations with the DOS approximation have been checked by a variational calculation using Gaussian trial wave functions for the $0^{--}$ states \cite{mathgauss}. Some items denoted by dots are dropped for the sake of clearness.}
\begin{tabular}{cllcccccc}
\hline\hline
  $K$ & $J^{PC} $ & $f_{J^P}$ & $M_G/\sqrt\sigma$ & \multicolumn{2}{c}{Model}  \\
&  &  & for $N_c\rightarrow\infty$ \cite{Teper} & DOS & Gaussian \\
\hline
 0 & $0^{--}$ & 1 & 5.91(25) & 6.55 &  6.49 \\
    2 & $0^{--*}$ & $\rho_+\rho_- + \lambda_+\lambda_-$ &7.63(37) &  8.77 & 8.53 \\
        & $0^{++}$ & $\rho_+\lambda_- - \rho_-\lambda_+$ & & 8.77 &  \\
       & $2^{--}$ & $\rho_+^2+\lambda_+^2+(+\longleftrightarrow -)$ & 7.89(35) & 9.02\\
      & $2^{+-}$ & $\rho_+^2+\lambda_+^2-(+\longleftrightarrow -)$ & 8.04(50) & 9.02\\
    3 & $1^{--}$ & $2\rho_+\rho_-\lambda_++\lambda_-(\rho_+^2-\lambda_+^2)+(+\longleftrightarrow -)$ &  9.36(60) & 9.74 \\   
      & $1^{+-}$ & $2\rho_+\rho_-\lambda_++\lambda_-(\rho_+^2-\lambda_+^2)-(+\longleftrightarrow -)$ &  9.43(75) & 9.74 \\   
        & $1^{++}$ & $-2\rho_+\lambda_+\lambda_-+\rho_-(\rho_+^2-\lambda_+^2)-(+\longleftrightarrow -)$ &  9.98(25) & 9.74 & \\   
      & $1^{-+}$ & $-2\rho_+\lambda_+\lambda_-+\rho_-(\rho_+^2-\lambda_+^2)+(+\longleftrightarrow -)$ &  10.06(40) & 9.74 &  \\   
       & $3^{\pm-}$ & \dots & & 10.09 & \\
         & $3^{\pm+}$ & \dots & & 10.09 & \\
     4 & $0^{--**}$ & $(\rho_+\rho_-+\lambda_+\lambda_-)^2$ & 8.96(65)  & 10.50 & 10.17 \\
      &  & $(\rho_+\lambda_-+\rho_-\lambda_+)^2 + (\rho_+\rho_--\lambda_+\lambda_-)^2$ &  & 10.50 &  \\
      & $0^{++*}$ & $(\rho_+\lambda_--\rho_-\lambda_+)(\rho_+\rho_-+\lambda_+\lambda_-) $& & 10.50 & \\
      & $2^{--*}$ & $(\rho_+\rho_-+\lambda_+\lambda_-)(\rho_+^2+\lambda_+^2)+(+\longleftrightarrow -) $&  9.46(66)& 10.63\\
       & $2^{+-*}$ & $(\rho_+\rho_-+\lambda_+\lambda_-)(\rho_+^2+\lambda_+^2)-(+\longleftrightarrow -) $&  9.97(91) & 10.63\\
       & $2^{\pm+}$ & \dots & & 10.63 & \\
       & $4^{\pm-}$ & \dots &  & 11.06 & \\
      5 & \dots & & & \\
      6 &   $0^{-+}$      &$(\rho_+\lambda_-+\rho_-\lambda_+)$  & 9.02(30) & 11.93  & \\
          &         &$\times\left[ 3 (\rho_+\rho_--\lambda_+\lambda_-)^2-(\rho_+\lambda_-+\rho_-\lambda_+)^2\right]$  &  &  & \\
              & \dots & & & &  \\
\hline\hline
\end{tabular}
\label{tab2}
\end{table}

\end{document}